\pdfoutput=1
\documentclass[10pt,conference]{IEEEtran}

\usepackage{enumitem} 
\newcommand{\bi}{\begin{itemize}[leftmargin=0.4cm]}
\newcommand{\ei}{\end{itemize}}
\newcommand{\be}{\begin{enumerate}[leftmargin=0.4cm]}
\newcommand{\ee}{\end{enumerate}}
\usepackage{balance}
\usepackage{hyperref}
\usepackage{times}
\usepackage{listings}
\usepackage{cite}
\usepackage{wrapfig,lipsum,booktabs}

\usepackage{pifont}
\newcommand{\cmark}{\ding{51}}%
\newcommand{\xmark}{\ding{55}}%

\newcommand{\tion}[1]{{Section }\ref{sect:#1}}

\usepackage{xcolor}
\definecolor{Gray}{gray}{0.85}
\usepackage{tikz}
\usepackage{framed}
\usepackage[framed]{ntheorem}

\usetikzlibrary{shadows}

\theoremclass{Lesson}
\theoremstyle{break}

\tikzstyle{thmbox} = [rectangle, rounded corners, draw=black, fill=Gray!40]

\newshadedtheorem{lesson}{Result}
\newshadedtheorem{lesson1}{Result}

\makeatletter
\let\th@plain\relax
\makeatother

\begin{document}
\pagestyle{plain} 


\title{Can You Explain That, Better? 
Comprehensible Text Analytics for SE Applications}

\author{\IEEEauthorblockN{Amritanshu Agrawal}
\IEEEauthorblockA{Computer Science Department\\
NC State, Raleigh, NC, USA\\
Email: aagrawa8@ncsu.edu}
\and
\IEEEauthorblockN{Huy Tu}
\IEEEauthorblockA{Computer Science Department\\
NC State, Raleigh, NC, USA\\
Email: hqtu@ncsu.edu}
\and
\IEEEauthorblockN{Tim Menzies}
\IEEEauthorblockA{Computer Science Department\\
NC State, Raleigh, NC, USA\\
Email: timm@ieee.org}}

\maketitle

\begin{abstract}
Text mining methods are used for a wide range of Software Engineering (SE) tasks. The biggest challenge of text mining is high dimensional data, i.e., a corpus of documents can contain $10^4$ to $10^6$ unique words. To address this complexity, some very convoluted text mining methods have been applied. 
Is that complexity necessary? Are there simpler ways to quickly generate models that perform as well as the more convoluted methods and also be human-readable?

To answer these  questions, we explore a combination of LDA (Latent Dirichlet Allocation) and  FFTs (Fast and Frugal Trees)  to classify NASA software bug reports from six different projects. Designed using principles from psychological science, FFTs return very small models that are   human-comprehensible.
When  compared to the commonly used text mining method and a recent state-of-the-art-system (search-based SE method that automatically tune  the control parameters of LDA), these   FFT models are very small  (a binary tree of depth $d=4$ that references only 4 topics) and hence easy to understand. They were also faster to generate and produced similar or better   severity predictions.

Hence we can conclude that,  at least for datasets explored here, convoluted text mining models can be deprecated in favor of  simpler method such as LDA+FFTs. At the very least, we recommend LDA+FFTs (a) when humans need to read, understand, and audit a model or (b) as  an initial baseline method for the SE researchers exploring text artifacts from software projects.

\end{abstract}


\section{Introduction}

Software analytics have been focusing on working with adept and state-of-the-art data miners in order to find the optimal results.
One sub-topic for software analytics is the use of sophisticated text mining techniques~\cite{tan1999text}. 
Text mining is much more complex task as it involves dealing
with high dimensional textual data that are inherently unstructured~\cite{zhang2008text,tan1999text}. 
These complex methods often generate   models not comprehensible to humans (e.g., using synthetic dimensions generated by an SVM kernel~\cite{menzies2009explanation}).
This complexity might not be necessary if simpler methods can be found to achieve the same performance,
while at the same time generating easy-to-understand models~\cite{vellido2012making}. We define our terminologies ``simple'' and ``comprehensible'' in this paper as:
\bi
\item \textbf{simple} - (1) has low dimensionality of features (in 10s and not 100 to 1000s); (2) generate small set of theories and (3) is not computationally expensive. Otherwise, we call it ``complex''.  
\item \textbf{comprehensible} - (1) comprise of small rules and (2) rules that quickly lead to decisions 
\ei

Moeyersoms et al.~\cite{moeyersoms2015comprehensible} comment that  predictive models not
only need to be accurate but also comprehensible, demanding that the user
can understand the motivation behind the model's prediction.  
They further remark that,
to obtain such predictive performance, comprehensibility is often sacrificed and
vice-versa. Do simpler methods perform worse? Martens et al.~\cite{martens2011performance} referred comprehensibility as to how well humans
grasp the classifier induced or how strong the mental fit of the classifier
is. 
Dejaeger et al.~\cite{dejaeger2013toward} said comprehensible models are often needed in order to inspire
confidence in a business setting and improve model acceptance. Business users are vocal in their complaints
about analytics~\cite{hihn2015data}, stating that there are rarely producible
models that business users can comprehend.

Researchers in SE use complex methods, such as Support Vector Machine (SVM) with 1,000 to 10,000s of Term Frequency (TF) or Term Frequency - Inverse Document Frequency (TFIDF) features in order to achieve high performance of prediction. Yet, they do not try to comprehend the model itself~\cite{lamkanfi2011comparing,xia2015automatic,kochhar2014automatic,xia2014automated,chaturvedi2012determining,sharma2012predicting} 
making business users more hesitant to adopt their methodologies and losing the value of their work.
Though, Latent Dirichlet Allocation (LDA) uses less number of features but does require 100s of features for finding an optimal model and be human-comprehensible~\cite{layman2016topic,chen2012explaining}. 
An alternative, better, search-based SE method (LDADE) was proposed recently which tries to find optimal parameters of LDA that can make the model more stable and achieve optimal results~\cite{agrawal2018wrong}. The problem with this model is that it is quite expensive in terms of CPU usage and still need 100s of features for it to be comprehensible. 
We need a simple method which: 
1) offers comparable performance; and
2) human comprehensible.


This paper study's  a simple data miner taken from the  psychological science literature, i.e., FFT which outputs small trees, (and generally, smaller is better comprehensible~\cite{brighton2006robust,gigerenzer1999good}).
In this study, FFT uses LDA features ($K=10$) with its default parameters, which does not require any expensive optimization to find the optimal K, and build its trees.
We seek few rules through FFT that can report severe and non-severe for the datasets under study. 
We compared this method against complex and most commonly used methods in SE literature, which are 1) TFIDF+SVM \cite{lamkanfi2011comparing, kochhar2014automatic, tian2015automated, tian2016unreliability}; and 2) a recent state-of-the-art system, LDADE+SVM~\cite{agrawal2018wrong}.
Based on this comparative analysis, we answer two research questions:


   \textbf{RQ1:  { How does simpler method \textit{perform} against most common sophisticated and recent state-of-the-art Search Based SE (SBSE) methods?}}


For software analytics, most text mining techniques use high dimensional TF or TFIDF features with complex classifiers like SVM~\cite{lamkanfi2011comparing,xia2015automatic,kochhar2014automatic,xia2014automated,chaturvedi2012determining,sharma2012predicting,tian2013drone,tian2015automated,thung2012automatic,tian2016unreliability,xia2017improving}. These features are large in number, in the range of 1,000 to 10,000s making any classifier, complex. 
Researchers shifted their focus on using LDA features in text mining since it is a good way for dimensionality reduction~\cite{blei2003latent,layman2016topic,zhou2016combining}. 
SBSE is recently introduced to find the optimal parameters at the expense of heavy runtime~\cite{agrawal2018wrong,agrawal2018better,fu2016tuning}. 
Agrawal et al.~\cite{agrawal2018wrong} tuned the parameters of LDA to find the optimal number of topics ($K$) which is further used by SVM for classification task (state-of-the-art SBSE method). 

We show that, FFT (with a depth, $d=4$) uses just 10 topics from LDA (simpler method) to achieve comparable performance as SVM with TFIDF features (sophisticated method) as well as LDADE with SVM (SBSE method). 
The runtime for building the simpler method is about 10 minutes slower than the sophisticated method's runtime but this may not be an arduous increase given the gains from its power of comprehensibility, whereas simpler method is 100 times faster than SBSE method.
Hence, we conclude that,



\begin{lesson}
Simpler method (LDA+FFT) offers similar performance as the sophisticated method (TFIDF+SVM) and the SBSE method (LDADE+SVM). Though simpler LDA+FFT method takes an extra 10  minutes than the baseline, but it is orders of magnitude faster than the SBSE method.

\end{lesson}

  \textbf{RQ2:  {Is simpler method more \textit{explainable or comprehensible} relative to the most common sophisticated and recent state-of-the-art SBSE methods?}}

We answered the question that simpler method can show comparable performance against sophisticated, and SBSE methods.
Now, we dive into the core of our study which is about comprehensibility. Why do we need comprehensible models? We need it to have some actionable insights from the model which will boost the confidence for businesses to accept the model for their software.

Representative characteristics help a model to be more explainable, i.e., small, visualized easily, and comprised of fewer rules that can quickly lead to decisions.
The range of features between 1,000 to 10,000s, makes any classifier big and non-comprehensible by default. 
LDA features offer more comprehensibility aspect to the model than TFIDF or TF features~\cite{zhou2016combining,layman2016topic}. 

We show that FFT with $K=10$ LDA features, referencing only 4 topics (depth, $d=4$) provide explainable model satisfying the characteristics mentioned earlier. 
Also,  we do not need a SBSE method which is orders of magnitude times slower to find optimal $K$, when a simpler method can provide a well comprehensible model. Hence, we conclude that

\begin{lesson}
FFT generates fewer rules referencing only 4 topics found by LDA are far more comprehensible than the most common sophisticated  and  SBSE methods.
\end{lesson}

\noindent
In summary, the main contributions of this paper are:
\bi
\item
A novel inter-disciplinary contribution of the application of psychological science in comprehensibility of text mining models.
\item LDA+FFTs offer comparable performance against a common text mining method, TFIDF+SVM.
\item LDA+FFTs are better, faster, and more comprehensible against the recent state-of-the-art method, LDADE+SVM.   
\item A new, very simple baseline data mining method (LDA+FFTs) against which more complex methods
can be compared.
\item
A reproduction package containing all the data and algorithms of this paper, see \url{https://github.com/ai-se/LDA_FFT}.
\ei

The rest of this paper is structured as follows:
\tion{background} talks about the background and theory of comprehensibility.
\tion{experiment} describes the experimental setup of this paper and above research questions are answered in
\tion{results}. Lastly, we discuss the validity of our results 
and a section describing our conclusions.

\section{Motivation and Related Work}
\label{sect:background}

This sections talks about theory of comprehensibility, the most commonly used text mining method for bug reports classification, curse of dimensionality, and power of computationally faster methods. We also show how FFTs are generated which is a great alternative to the existing approaches.

\subsection{Theory of Comprehensibility}
\label{sect:comprehensibility}


For software analytics, it is a necessity to find such models that can produce simple and actionable insights for the software practitioners to interpret and act upon~\cite{chen2018applications}.
Models
are effectively useless if they cannot be interpreted by researchers, developers, and testers~\cite{vellido2012making}.
Business users have been vocal in their complaints
about analytics~\cite{hihn2015data}, saying that there are rarely producible
models that they can comprehend.
According to several researchers \cite{kimicse16, explainableicse18, lipton2016}, actionable insights from software artifacts are the core deliverable of software analytics. 
These insights are then used by the users to enhance their productivity, which is measured in terms of the task that are accomplished. However, is model comprehensibility taken into consideration in the process of development?


Machine learners generate theories and people read theories. 
But how many of such learners generate the kind of theories that  machine learning practitioners can read? 
In practice, with availability of big data and tremendous amount of information, yet limited time and resources to explore, such as manager  rushing with deadlines to release a software or stockbrokers making instant decisions about buying or selling stocks. 
Rather, in such a critical situation, a person might instead just want to have the least expert-level comprehension of that domain to  achieve the most benefits. It therefore follows that machine learning for these practical cases should not strive for elaborated theories or expressive power of the language. 
A better goal for machine learning would be to find the smallest set of theories with the most impacts and benefits~\cite{menzies2003data}.




Also, in today's businesses, the problem is
not accessing data but ignoring the irrelevant
data. Most modern businesses can electronically access large amounts of data such
as transactions for the past two years or
the state of their assembly line. The trick is effectively using the available data. In practice, this means summarizing large datasets to find the ``pearls in
the dust'' - that is, the data that really matters~\cite{menzies2003data}.

That is why, Gleicher~\cite{gleicher2016framework} developed their
 framework of comprehensibility~\cite{gleicher2016framework} and concluded that many researchers do not consider the power of comprehensibility and miss out on important aspects of their results.
According to Gleicher:
\be
\item Comprehensibility makes us understand a prediction to appropriately \textit{trust} it, or a predictive process to trust in its
ability to make predictions.

\item Comprehensibility helps in \textit{prescriptiveness}, which is the quality of a model that allows its user to act on something with a result, e.g., its ability to inform action.

\item Understanding of a model can drive iterative refinement that is applied to
\textit{improve} predictive accuracy, efficiency, and robustness.

\item While a statistical model usually uncovers correlations,
\textit{discovers causality}, it can also be a useful starting point for \textit{theory building}, or  an
 approach towards \textit{testing theory}.
\item Comprehensibility can \textit{characterize} by easily interpreting what the model can do and where it can be applied.

\item It can \textit{generalize} modeling to other situations which can be part of other (future) applications.

\item It can identify the success (or failures) in one model, modeling application, or modeling
process, that can help us to \textit{improve our practices} for future applications.

\ee

Comprehensibility is defined as
the ability of the various stakeholders to understand relevant aspects of the modeling process. How can a model be comprehensible? According to various researchers~\cite{gleicher2016framework,vellido2012making,martens2007comprehensible,martens2013explaining}, a comprehensible model can be represented with a rule-based learning~\cite{phillips2017fftrees, brighton2006}, or size of the output, i.e., smaller models~\cite{maimon2005decomposition}, or better visualization~\cite{gleicher2016framework}.

\begin{figure}[!t]
        \centering
        \includegraphics[width=\linewidth]{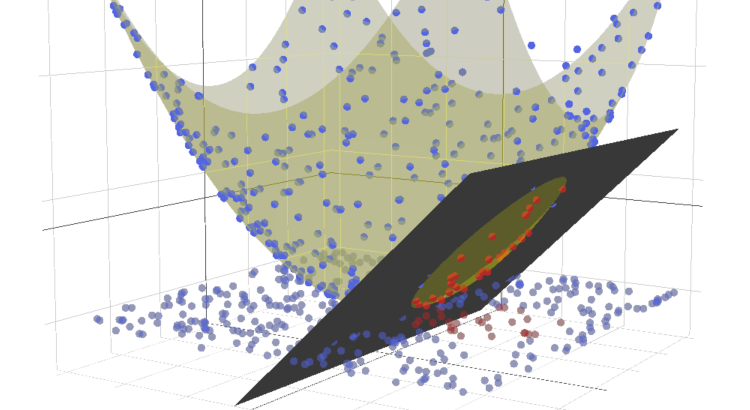}
    \caption{ An example of SVM model from ~\cite{svmmodel}  }
    \label{fig:svm_model}
    \vspace{-0.5cm}
\end{figure}

According to Phillips et al.~\cite{phillips2017fftrees}, a model shown to be comprehensible enough for human, when a human can fit the model into their Long Term Memory (LTM)~\cite{larkin1980expert} and when the rules within the model can efficiently lead to decisions. Imagine a model as shown in Figure~\ref{fig:svm_model} of SVM, a human would not be able to reason from such a sophisticated output because of 2 reasons: 
1) The model is mostly points of transformed data on a new multi-dimensional feature space automatically inferred by some kernel function. Due to the arcane nature of these kernels, it is hard for humans to attribute meaning to these points~\cite{nasrabadi2007pattern,menzies2009explanation}; and
2) The model infers a decision boundary or hyperplane (as shown in Figure~\ref{fig:svm_model}) without any generalization~\cite{haasdonk2005feature}.  A SVM defines its decision boundary in terms of the vectors nearest that boundary. All these ``support vectors'' are just points in space so understanding any one of them incurs the problems.

\begin{figure}[!b]
\vspace{-0.5cm}
 {\normalsize
\begin{minipage}{\linewidth}
\begin{tabular}{p{\linewidth}}
  \qquad  if  \qquad \ \ topic 1 $>$  0.80   \qquad \qquad  then false \\
  \qquad  else if \ \ topic 7 $>$  0.60    \qquad \qquad   then true \\
   \qquad else if \ \ topic 3 $>$  0.65   \qquad \qquad  then true \\
   \qquad else if \ \ topic 5 $\leq$ 0.50    \qquad \qquad  then true \\
   \qquad else \qquad \qquad \qquad \qquad \qquad \qquad false \\
\end{tabular}
\end{minipage}
}
\caption{Example of a much simpler FFT model. How this FFT is generated is explained in \tion{fft}. The premise of this paper is that
such simple models can perform as well, or better than more complex models that use extra dimensions, like Figure~\ref{fig:svm_model}. This is an example created by our proposed method on the dataset PitsA under study.}\label{fig:fft}
\end{figure}

Further, SVMs offer much less support for understanding the entire set of these points than, say, some rule-based representation (as shown in Figure~\ref{fig:fft} which is an example created by our proposed method on the dataset under study). 
To understand this, consider a condition $a>4$ that might be found in a rule-based representation, and within the hyperspace of all data, this inequality defines a region within which center conclusions are true, regardless of other attributes. That is, this condition $a>4$ is a generalization across a large space of examples, a region that humans can understand as ``within this space, certain properties exist''.
The same is not true for support vectors. Such vectors do not tell humans which attributes are most important for selecting one conclusion over another, nor can they divide a space of examples into multiple regions. Rule-based representations do not have that limitation. They can divide space into multiple sectors within which humans know how far they can adjust a few key attributes in order to move from one classification to another.


Consequently, psychological scientists have developed FFT as a rule-based model that is quickly comprehensible, comprising of few rules. A FFT tree is a binary tree classifier, where either one or both node has a terminating  branch to a decision node. Basically, it will trigger an immediate understanding and action for each question being asked or topic information feature. As shown in Figure~\ref{fig:fft}, the same complex model of Figure~\ref{fig:svm_model} can be comprehensible enough using FFT which is just 5 lines of rules. We will study FFT in greater detail, later in the \tion{fft}.

Menzies et al.~\cite{menzies2008automated} obtained similar Decision Tree (DT) rules for the same dataset PitsA which is under study in this paper. 
A condensed example of their rules are shown in Figure~\ref{fig:dttim}, the conditions in these rules are at the term occurrence level, whereas our example of FFT (Figure~\ref{fig:fft}) are at topic information level. 
The term occurrence condition failed to provide any generalized intuition or expert comprehension of how to use such a rule to classify bug report automatically. 
But if we consider our proposed FFT tree, we observed that {\em if topic 3 $>$ 0.65 then the report can be classified as severe}. 
The top terms denoting topic 3 are \textbf{messag unsign bit code file byte word ptr} and we can say these terms generalizing ``type conversion'' topic. 

Developers can now use this information to avoid future mistakes in the code where type conversion is happening. We contacted the original
users of the PITS data~\cite{menzies2008automated} to look at the topics which we generated (and the conditions where  they were found). 
They agreed that their rules were not \textit{generalizable};
i.e. they could not use those rules to \textit{improve} their systems but the topics which we generated are highly relevant and practical. 
This validates and motivates that the rules generated by our FFT on a topic occurrence level are more comprehensible.

\begin{figure}[!t]
 {\small
~~~~~\begin{minipage}{\linewidth}
\begin{tabular}{p{\linewidth}}
  \quad \quad if \ sr $\leq$  0 \& rvm $\geq$  1 \& l4 $\geq$ 1 \& cdf $\geq$ 1 \ \ \ \ then 4 \\
  else if \ sr $\leq$  1 \& issu $\geq$  1 \& code $\geq$ 3  \qquad  \qquad \ \ \  then 4 \\
  else if \ control $\geq$  1 \& code $\geq$ 1 \& attitud $\geq$ 4  \qquad  then 2 \\
  else if \ l3 $\geq$  2  \& obc $\leq$ 0 \& perform $\leq$ 0 \qquad \qquad  then 2 \\
  else if \ script $\geq$ 1 \& trace $\geq$ 1    \qquad \qquad \qquad  \qquad \ \ then 3 \\
  else \qquad \qquad \qquad \qquad \qquad \qquad \qquad \qquad \qquad \qquad \qquad 3
\end{tabular}
\end{minipage}
}
\caption{Similar Decision Tree (DT) rules obtained by Menzies et al.~\cite{menzies2008automated}  for PitsA (the dataset under study in this paper).}
\label{fig:dttim}
\vspace{-0.5cm}
\end{figure}


While this paper places high value on comprehensibility, we note that much prior work has ignored this issue.
In March 2018,  we searched Google scholar for the papers that are published in the last decade, which does text mining to build defect/bug predictors and also talks about comprehensibility. 
From that list, we selected ``highly-cited'' papers, which we defined as having more than 5 citations per year. 
After reading through the titles and abstracts of those papers, and skimming the contents of the potentially interesting papers, we found 16 papers as shown in Table~\ref{tbl:survey} that motivates our study.

From Table~\ref{tbl:survey}, we can see that despite the importance of method comprehensibility as pointed out by Gleicher~\cite{gleicher2016framework} and others,  all the 16 ``highly-cited'' papers talk about comprehensibility in some form but do not have few rules which are browsable and can fit into human's LTM.

\subsection{Bug Reports Classification}
\label{sect:bug}

The case studies used in this paper comes from text classification of bug reports.
This section describes those case studies.

Many SE text mining researches have been done on bug reports classification to categorize the description of the fault occurrence in a software system. 
 Zhou et al.~\cite{zhou2016combining} found the top 20, 50, 100 top terms and used these as features to model Naive Bayes, and Logistic regression classifiers. They reported on precision, recall and f-score, and concluded that their method had a significant improvement over other proposed methods. Yet, they did not use these top terms to comprehend the prediction model. 
 Menzies et al.~\cite{menzies2008automated} used TFIDF featurization technique with Naive Bayes classifier to predict the severity of defect reports and they lacked in showing how to interpret such a method. 
 Few researchers~\cite{chaturvedi2012determining,sharma2012predicting} used only top TF features to build a SVM classifier but did not provide interpretability of the method.

  \begin{table}[!t]
\centering
\caption{Highly Cited Papers}
\label{tbl:survey}
\resizebox{\linewidth}{!}{
    \begin{tabular}{c|c|c|c|c|c}
        \begin{tabular}[c]{@{}c@{}}\textbf{Ref}\end{tabular} & \textbf{Year} & \textbf{Citations} & \begin{tabular}[c]{@{}c@{}}\textbf{SVM as} \\\textbf{a Classifier} \end{tabular} &\begin{tabular}[c]{@{}c@{}} \textbf{High} \\\textbf{Dimensional} \\\textbf{Features}\end{tabular}&\begin{tabular}[c]{@{}c@{}} \textbf{Comprehensibility} \end{tabular}\\ \hline
        \cite{menzies2008automated} & 2008 & 191 & \xmark & \cmark & \xmark \\
        \cite{lamkanfi2011comparing} & 2011 & 119 & \cmark & \cmark & \xmark \\
        \cite{thung2012automatic} & 2012 & 66 & \cmark & \cmark & \xmark \\
        \cite{zhou2016combining} & 2016 & 51 & \xmark & \cmark & \xmark \\
        \cite{tian2013drone} & 2013 & 50 & \cmark & \cmark & \xmark \\
        \cite{chen2012explaining} & 2012 & 47 & \xmark & \cmark & \xmark \\
        \cite{chaturvedi2012determining} & 2012 & 38 & \cmark & \cmark & \xmark \\
        \cite{xia2015automatic} & 2015 & 28 & \cmark & \cmark & \xmark \\
        \cite{pingclasai2013classifying} & 2013 & 25 & \xmark & \cmark & \xmark \\
        \cite{sharma2012predicting} & 2012 & 25 & \cmark & \cmark & \xmark \\
        \cite{xia2014automated} & 2014 & 22 & \cmark & \cmark & \xmark \\
        \cite{tian2015automated} & 2015 & 20 & \cmark & \cmark & \xmark \\
        \cite{xia2017improving} & 2017 & 16 & \cmark & \cmark & \xmark \\ 
        \cite{kochhar2014automatic} & 2014 & 14 & \cmark & \cmark & \xmark \\  
        \cite{tian2016unreliability} & 2016 & 8 & \cmark & \cmark & \xmark \\
        \cite{layman2016topic} & 2016 & 6 & \xmark & \cmark & \xmark \\
\end{tabular}}
\vspace{-0.4cm}
\end{table}

Many other researchers used SVM as a classifier but used high number of TF features to do bug/defect prediction ~\cite{lamkanfi2011comparing,xia2015automatic,kochhar2014automatic,xia2014automated} and they provided top significant terms to explain about the cause of these bugs. In other works, few researchers used SVM with high number of TF features but did not report terms to provide any explanation~\cite{tian2013drone,tian2015automated,thung2012automatic,tian2016unreliability}.

Researchers also used LDA's document topic distribution as features to build bug report prediction models~\cite{pingclasai2013classifying,layman2016topic,chen2012explaining,xia2017improving}.
Xia et al.~\cite{xia2017improving} worked on LDA features with SVM classifier but did not have any interpretability power. Pingclasai et al.~\cite{pingclasai2013classifying} compared different size of topics needed by LDA against different number of top TF features. They found that LDA with $K=50$ yields the best f-score. 
Layman et al.~\cite{layman2016topic} used different number of topics to identify severity of bug reports on 6 NASA Space System Problem datasets. 
They also comprehensibly showed what these reports were talking about. The problem with this was that they chose high number of topics. 
Also, Chen et al.~\cite{chen2012explaining} used LDA to identify whether defect prone module stays defect prone even in future versions. They showed top topics with top words related to defect. 
But the problem existed similar to Layman et al., that they used high number of topics.

We looked at recent studies, which uses high dimensional features combined with different classifiers such as Naive Bayes, SVM, Logistic regression~\cite{lamkanfi2011comparing,zhou2016combining,menzies2008automated} to accurately model the data. 
But out of that, SVM is the most commonly, frequently and popularly used classifier. From Table~\ref{tbl:survey}, we can see that 11/16 (about 70\%) highly cited papers used SVM as classifiers. Therefore, we chose SVM classifier as the complex baseline learner to compare against the simple FFT model.  

\subsection{Curse of Dimensionality}
\label{sect:dimension}

All the text mining techniques model high dimensional data, i.e., a corpus of documents that contains $10^4$ to $10^6$ unique words. 
The common problem associated with such data is that when the dimensionality increases, the volume of the space increases drastically which leads to available data getting sparsed~\cite{nasrabadi2007pattern}. 
This sparsity is problematic when we  try to find statistically sound and reliable result, the amount of data needed to support the result often grows exponentially with the dimensionality. 
Also, modeling such high dimensional data often relies on detecting areas where objects form groups with similar properties, however in high dimensional data, all objects appear to be sparse and dissimilar in many ways, which prevents common data organization strategies from being efficient~\cite{friedman1997bias}.

High dimensional data also increases the complexity for data modeling, and is a curse for finding comprehensible models. 
Researchers use TF and TFIDF feature extraction techniques~\cite{menzies2008automated,xia2014automated} which provides 1,000 to 10,000s of features for a learner to model it. 
These numerous features would not offer smaller concise comprehensible models. From Table~\ref{tbl:survey}, we can see that all the 16 papers have high dimensional features driving us to find alternate methods for reduction in dimensionality.

To tackle the curse of dimensionality, researchers employ different dimensionality reduction techniques like feature transformation (Principal Component Analysis, Latent Dirichlet Allocation), Sampling, Feature Selection Techniques, and many more~\cite{van2009dimensionality,fodor2002survey}. 
For text mining, researchers used mostly a feature transformation or feature selection technique to reduce the feature space in order to find the top words from the corpus which can the be used in classifiers~\cite{zhou2016combining,chaturvedi2012determining,sharma2012predicting}.

Latent Dirichlet Allocation (LDA) is a common technique observed in text mining for dimensionality reduction~\cite{layman2016topic,agrawal2018wrong}. 
LDA provides topics that are comprehensible enough and researchers can browse through them to make decisions as shown by Agrawal et al~\cite{agrawal2018wrong}. We agree with their work and their motivation of choosing such a feature extraction technique. 
That's why, we chose LDA as a feature extraction technique (since we get concise topics) and after combining it with FFT (depth, $d=4$), we get few rules that are comprehensible enough while having better or comparable results classification performance.

\subsection{Computationally Inexpensiveness}
\label{sect:computation}

There always exists a trade-off between the effectiveness and the cost of running any method.
The method should not be expensive to apply (measured in terms
of required CPU, or runtime).
Before a community can adopt
a method, we need to first ensure that the method
executes very quickly. Some methods, especially which are used to solve the problem of hyperparameter optimization (the problem of choosing a set of optimal parameters for a learning algorithm),
can require hours to days to years of CPU-time to
terminate~\cite{wang2013searching,agrawal2018wrong}. Hence, unlike such methods, we need to
 select baseline methods that are reasonably fast.
 
 One such resource expensive method is recently introduced by Agrawal et al.~\cite{agrawal2018wrong}, where they optimized the hyperparameters of LDA to find the optimal settings. 
 They optimized the LDA for $\Re_n$ score which was the measure of how stable the generated topics are. 
 They showed that stable topics are needed if developers/users are using these topics for further analysis, especially when it comes to unsupervised learning.
 They also used these stable topics for supervised learning and showed that the prediction performance is comparable against the commonly used text mining technique of TFIDF with SVM classifier. 
 The major drawback with their method is that it is computationally expensive, and is about three to five times slower.
 It is computationally expensive due to 2 reasons: 1) Use of computationally expensive optimizer (Differential Evolution) and 2) Number of Topics, which has direct relation with its runtime, i.e., the more number of topics, the more the run time.
 
As previously mentioned, the reason for choosing LDA features was its power of comprehensibility. Though we do not want to use  an expensive technique like LDADE, when we have the option of using default parameters without sacrificing the performance while achieving much better comprehensibility with FFT.

\subsection{How are FFTs  generated?}
\label{sect:fft}

Psychological scientists have developed FFTs (Fast and Frugal Trees) as one way
  to generate comprehensible models consisting of
   separate tiny rules~\cite{phillips2017fftrees,chen2018applications,martignon2008categorization}.
      A FFT is a decision tree made for binary classification problem with exactly two branches extending
        from each node, where either one or both branches is an exit
        branch leading to a leaf~\cite{martignon2008categorization}. 
        That is to say, in an FFT,  every question posed by a node will
        trigger an immediate decision
        (so humans can read every leaf node
        as a separate rule).

We used the similar implementation of FFT as offered by Fu and Chen et al.~\cite{fu2018building,chen2018applications}. 
An FFT of depth $d$ has a choice of two ``exit policies'' at each level: the existing branch can select for the negation of the target, i.e., non-severe, (denoted ``0'') or the target (denoted ``1''), i.e., severe.
The right-hand-side tree in Figure~\ref{fig:fft1} is 01110 since:
\bi
\item
The first level found a rule that exits to the negation of the target: hence, ``0''.
\item
While the next tree levels found rules that
exit first to target; hence, ``111''.
\item
And the final line of the model exits
to the opposite of the penultimate line; hence, the final ``0''.
\ei

\begin{figure}[!b]

 {\normalsize
\begin{minipage}{\linewidth}
\begin{tabular}{p{\linewidth}}
  \qquad  if  \qquad \ \ topic 1 $>$  0.80   \qquad \qquad  then false \qquad \#0 \\
  \qquad  else if \ \ topic 7 $>$  0.60    \qquad \qquad   then true \qquad \  \#1 \\
   \qquad else if \ \ topic 3 $>$  0.65   \qquad \qquad  then true \qquad \  \#1 \\
   \qquad else if \ \ topic 5 $\leq$ 0.50    \qquad \qquad  then true \qquad \  \#1 \\
   \qquad else \qquad \qquad \qquad \qquad \qquad \qquad false \qquad \qquad \ \#0 
\end{tabular}
\end{minipage}
}
\vspace{-0.25cm}
\caption{Example of an FFT}
\label{fig:fft1}
\end{figure}

Following the advice of~\cite{fu2018building,chen2018applications,phillips2017fftrees}, for all the experiments of this paper, we use a depth    $d=4$. 
For trees of depth $d=4$, there are $2^4=16$ possible trees which can be denoted as 00001, 00010, 00101, ... , 11110. During FFT training, all $2^d$ trees are generated, then we select the best one (using the training data).
 This single best tree is then applied to the test data.
 Note that FFTs of such small
depths are very succinct
(see examples in Figures~\ref{fig:fft} and~\ref{fig:fft1}). Such FFTs generate rules which leads to decision of finding a report as severe and non-severe for the datasets under study. Many other data mining algorithms used in software analytics are far less
succinct and far less comprehensible as explained in \tion{comprehensibility}.

\section{Experimentation}
\label{sect:experiment}

All our data, experiments, scripts are available to be downloaded from \url{https://github.com/ai-se/LDA_FFT}.

\subsection{Dataset}


\textbf{PITS} is a widely used text mining dataset in SE studies~\cite{menzies2008automated,menzies2008improving,layman2016topic}. The dataset is generated from NASA software project
and issue tracking system (PITS) reports~\cite{menzies2008improving, menzies2008automated}. This text discusses
bugs and changes found in big reports and  review patches.
Such issues are used
to manage quality assurance, to support communication
between developers. Text Mining techniques can be used
to predict each severity separately~\cite{layman2016topic}. The dataset can be downloaded from \url{http://tiny.cc/seacraft}. Note that, this data comes from six different
NASA projects, which we label as PitsA, PitsB, and so on. For this study, we converted these severity into binary classification where the max number of reports with one severity is labeled as positive class and the rest as negative. 
We employed the usual preprocessing steps mentioned in text mining literature~\cite{agrawal2018wrong, feldman2006tmh} which are tokenization, stop-words removal, and stemming. Table~\ref{tb:dataset} shows the number of documents, feature size, and the percentage of severe classes after preprocessing. 

 \begin{table}[!t]
\begin{center}
\caption{Dataset statistics. Data comes from the SEACRAFT repository: \url{http://tiny.cc/seacraft}}
\label{tb:dataset}
\begin{tabular}{c@{~}|r@{~}|r@{~}|r@{~}}
\begin{tabular}[c]{@{}c@{}} \textbf{Dataset} \end{tabular} & \begin{tabular}[c]{@{}c@{}} \textbf{No. of Documents}\end{tabular} & \textbf{Feature Size} & \begin{tabular}[c]{@{}c@{}} \textbf{Severe \%}\end{tabular} \\ \hline
PitsA & 965 & 2001  & 39  \\ 
PitsB &   1650 & 1685  & 40  \\  
PitsC &   323 & 544  & 56 \\ 
PitsD &   182 & 557 & 92  \\  
PitsE & 825 & 1628  & 63 \\ 
PitsF & 744 & 1431 & 64 \\ 

\end{tabular}
\end{center} 
\vspace{-0.5cm}
\end{table}

\subsection{Feature Extraction}

Textual data are actually series of words. In order to run machine learning algorithms we need to convert the text into numerical feature vectors. We used 2 types of feature extraction techniques:

\bi
\item \textit{Term Frequency-Inverse Document Frequency (TFIDF)}: If a word occurs $w$ times
  and is found in $d$ documents  and there
  are $W$, and $D$ as total number of words and documents respectively~\cite{agrawal2018wrong}, then TFIDF is scored
  as follows:
  \[
  \mathit{TFIDF}(w,d)=   \frac{w}{W} *\log{\frac{D}{d}}\]

\item \textit{Topic Information Features (LDA)}: We need to decide the number of topics size before applying the
LDA model to generate topic information features. To identify the number of topics we employed 2 strategies: 1) Manual number of topic size (10, 25, 50, 100) and 2) Choosing an optimal K using LDADE method~\cite{agrawal2018wrong}. LDA model produces the probability of a document in each topic which is used as a feature vector. Normally, the number of topics is significantly smaller than
the number of terms, thus LDA can effectively reduce the feature dimension~\cite{blei2003latent}.
\ei

\begin{figure*}[!t]
\centering
        \includegraphics[width=\textwidth ]{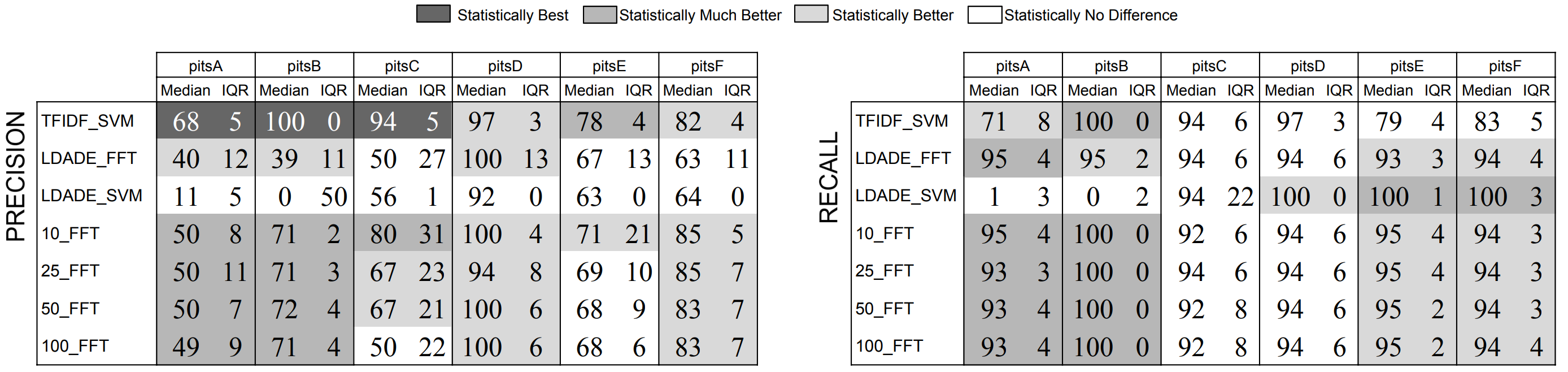}
    \caption{Comparison of LDA (K=10, 25, 50, 100) with FFT against TFIDF+SVM, LDADE+SVM and LDADE+FFT. Columns represent different datasets under study and scored on precision and recall. We show median and IQR (inter-quartile range, 75th-25th percentile) values. Different color coding shows the results from Scott-Knott procedure. The statistical comparison is across rows to find which method performs the best.}
    \label{fig:stats}
    \vspace{-0.5cm}
\end{figure*}

\subsection{Classifier}

For this study we used 2 machine learning algorithms, 1) \textit{Support Vector Machine (SVM)} and 2) \textit{Fast and Frugal Trees (FFTs)}. We use these, as explained earlier in \tion{background}. Though, there are other available choices like Deep learning, Decision Tree (DT), and Random Forest (RF) which have shown to be powerful in SE applications~\cite{ghotra2015revisiting, agrawal2018better,white2015toward,yang2015deep}. However, deep learning does not
readily support explainability, they have been criticized as ``data
mining alchemy''~\cite{DL2017alchemy} and also a recent study by Majumder et al.~\cite{majumder2018} suggest it may not be the most useful for SE data. DT or RF can generate small set of rules but performance can be sacrificed. Camilleri et al. \cite{Camilleri2014DT} showed that, DT have $70\%$ accuracy and significantly increased to $84\%$ when the depth of the tree increased from 0 to $10$, meaning that rules generated also moved from less to many. Hence, DT or RF may not be useful for this study.

Using a dataset, a performance measure and a classifier, this experiment conducts the 5*5 stratified~\cite{refaeilzadeh2009cross,kohavi1995study} cross-validation study to make our results more robust and reliable. This checks the amount of variance for such learners. The variance should be as minimal as possible. To control the randomization, seed is set so that the results can be reproducible. For implementation  of SVM and other methods, we used  the open source tool Scikit-Learn \cite{pedregosa2011scikit} and we relied upon their default parameters as our baseline.
Our stratified cross-validation study~\cite{refaeilzadeh2009cross, agrawal2018better} which includes the process of DE is defined as follows:
\bi
\item We randomized the order of the dataset set five times. This reduces the sampling bias, that some random ordering of examples in the data can conflate our results.
\item Each time, we divided the data into ten bins.
\item For each bin (the test), we trained on four bins (the rest) and then tested
on the test bin.
\item
When using LDADE, we further divide those four bins of training data. three bins are used for training the model, and one bin is used for validation in DE. DE is  run to improve the performance measure when the LDA was applied to the training data.
Important point:  When tuning, this rig \underline{{\em never}} uses test data.
\item
The model is applied to the test data to collect scores. 
\ei

\subsection{Evaluation Measure}

The problem studied in this paper is a binary classification task.
The performance of a binary classifier can be assessed via a  confusion matrix as shown in Table~\ref{fig:cmatrix}
where a ``positive'' output is the positive class  under study and a ``negative'' output is the negative one.

\begin{wraptable}{r}{1.5in}
\small
\begin{center}
\vspace{-1cm}
\caption{Results Matrix}
\label{fig:cmatrix}
\begin{tabular} {@{}cc|c|c|l@{}}
\cline{3-4}
& & \multicolumn{2}{ c| }{Actual} \\ \cline{2-4}
& \multicolumn{1}{ |c| }{Prediction} & false & true  \\ \cline{2-4}
& \multicolumn{1}{ |c| }{negative} & $\mathit{TN}$ & $\mathit{FN}$ & \\ \cline{2-4}
 & \multicolumn{1}{ |c| }{positive} & $\mathit{FP}$& $\mathit{TP}$  &  \\ \cline{2-4}
\cline{2-4}
\end{tabular}
\end{center}
\vspace{-0.5cm}
\end{wraptable}

Further, ``false'' means the learner got it wrong and ``true'' means the learner correctly identified
a positive or negative class. Hence, Table~\ref{fig:cmatrix} has four quadrants containing, e.g., $\mathit{FP}$ which denotes ``false positive''.

We used the following 2 measures that can be defined from this matrix as: 
\bi
\item \textbf{Recall} $=$ $pd$  $=$ $\mathit{TP}/(\mathit{TP} + \mathit{FN})$

\item  \textbf{Precision}  $=$ $prec$ $=$ $\mathit{TP}/(\mathit{TP} + \mathit{FP})$
\ei

No evaluation criteria is ``best'' since different criteria are appropriate
in different real-world contexts. Specifically, in order to optimize the performance of the released software, management would maximize the precision which would reduce the recall. 
When dealing with safety-critical applications, management may be ``risk adverse'' 
and hence many elect to maximize recall, regardless of the time wasted exploring false alarm~\cite{agrawal2018better}. Both precision and recall cannot be maximized at the same time. We assume that this holds true in the context of this paper and a business user wants to maximize either precision or recall and that is why we evaluate FFT on individual scores.

\subsection{Statistical Analysis}
\label{sec:scott-knott}

We compared our results using statistical significance test and an effect size test. Significance test is useful for detecting if two populations differ merely by random noise. Scott-Knott procedure was used as significance test~\cite{mittas2013ranking,ghotra2015revisiting,amrit:icse:seip2018}.

Effect sizes are useful for checking whether two populations differ by more than just a trivial amount. A12 effect size test was used~\cite{arcuri2011practical}. Our stats test are statistically significant with 95\% confidence and not a ``small'' effect ($A12 \ge 0.6$).

\section{Results}
\label{sect:results}

\textbf{RQ1: {How does simpler method perform against most common sophisticated  and recent state-of-the-art Search Based SE (SBSE) methods?} }

As discussed in \tion{bug}, we found that the most common text mining technique for binary classification in software engineering is TFIDF as the feature extraction method with SVM as a classifier.
In recent studies~\cite{agrawal2018wrong,layman2016topic}, LDA feature extraction is shown to be of a great alternative due to it achieving similar performance as well as reduction in dimensionality. 

Some researchers also adapted hyperparameter tuning to optimize performance but they do come with an expense of heavy runtime~\cite{fu2016tuning,ghotra2015revisiting,agrawal2018wrong,agrawal2018better}. 
Agrawal et al.~\cite{agrawal2018wrong} showed LDADE with SVM (SBSE method) to achieve better performance for classification tasks. LDADE finds optimal $K$, $\alpha$ and $\beta$, but $K$ matters the most for supervised learning~\cite{agrawal2018wrong}.

FFT is shown to be a good classifier when dealing with low dimensionality in defect prediction studies~\cite{fu2018building,chen2018applications}. We used LDA as features for FFT due to its power to explain about the text.
That is why we compared sophisticated method (TFIDF+SVM) as well as SBSE method (LDADE+SVM) against the proposed simpler method (LDA+FFT). 
We also compared LDADE+FFT against LDA+FFT, and  tried with different variants of FFTs by using  different topic sizes ($K=10, 25, 50, 100$), changing K manually rather than using an automatic technique like LDADE which is an expensive task, to see what improvement can we find.

Figure~\ref{fig:stats} offers a statistical analysis of different results achieved between TFIDF+SVM, LDADE+SVM, LDADE+FFT against 10\_FFT, 25\_FFT, 50\_FFT, 100\_FFT. Each column represents different datasets and each sub-figure shows  precision and recall scores. We assume that business users want to maximize either precision or recall and that is why we run FFTs separately on individual scores.
We report median and IQR (inter-quartile range, 75th-25th percentile) values, and \textit{darker} the cell, the statistically \textit{better} the performance. For example, in sub-figure where we report precision values, consider the column of pitsA dataset, we will read across rows to know which method works the best. In this case, TFIDF\_SVM is better across other methods. Similarly other dataset's results can be read.
Also, if the same color exists across, they are either statistically insignificant or are different only via a \textit{small} effect (as stated by the statistical methods described in Section \ref{sec:scott-knott}).  

For recall, we observe that 10\_FFT, 25\_FFT, 50\_FFT, and 100\_FFT (LDA\_FFTs) are performing statistically similar against all 6 datasets, 
whereas for precision scores, 10\_FFT, 25\_FFT, 50\_FFT, 100\_FFT are performing similar in 4 out of 6 datasets and 10\_FFT wins on the remaining 2 occasions. 
This came as a surprise since value of K are shown to have effect on the classification performance in recent SBSE method~\cite{agrawal2018wrong} whereas FFT has minimal effect on what value of $K$ is used. From now on, that is why all our comparisons are with 10\_FFT.

We note that simpler methods (10\_FFT) are statistically better or similar on 5 out of 6 datasets against TFIDF+SVM (sophisticated method) when compared on recall but it  performs similar on 2 out of 6 datasets when we look at precision value.
This tells that simple FFT method have comparable performance against the complex method.

We also found that 10\_FFT is winning on precision by a big margin on all 6 datasets when compared against LDADE\_SVM. On the other hand, 10\_FFT method offered comparable performance against the other 6 datasets for recall.
This changes a recent study's conclusion~\cite{agrawal2018wrong} where Agrawal et al. showed LDADE\_SVM, new simpler state-of-the-art method, defeating the sophisticated method (TFIDF+SVM). The datasets under study are different than what Agrawal et al. used, which might have affected our results.
Though, our findings say that:

\begin{center}
\vspace{2mm}
{\bf {\em LDADE+SVM is worse than LDA+FFT and TFIDF+SVM but LDA+FFT is similar to TFIDF+SVM.}}
\end{center}

\bi
    \item LDA\_FFT with $K=10$ offers comparable performance against TFIDF+SVM.
    \item LDA\_FFT with $K=10$ are wining against LDADE+SVM in majority cases.
\ei
\vspace{2mm}



With any empirical study, besides classification power, we have to look at the runtimes as another criteria to evaluate the methods performance.
Table~\ref{tbl:runtimes} shows the runtimes in minutes. 
From the table, it can be observed that LDA+FFT is only somewhat slower than TFIDF+SVM which may not be an  arduous  increase  given  the  gains  from  its  power  of comprehensibility discussed in \textbf{\textit{RQ2}}.
However, it can be observed that LDA+FFT combination is orders of magnitude faster (100 fold) than SBSE method (LDADE+SVM). 
This concludes that SBSE method is quite expensive and our picked alternative solution, i.e., LDA+FFT, is a promising candidate. 

 \begin{table}[!t]
\begin{center}
\caption{Runtimes (in minutes)}
\label{tbl:runtimes}
\begin{tabular}{c@{~}|r@{~}|r@{~}|r@{~}}
\begin{tabular}[c]{@{}c@{}} \textbf{Dataset} \end{tabular} & \begin{tabular}[c]{@{}c@{}} \textbf{TFIDF\_SVM}\end{tabular} & \textbf{10\_FFT} & \begin{tabular}[c]{@{}c@{}} \textbf{LDADE\_SVM}\end{tabular} \\ \hline
PitsA & $<$1                &   $<$8           & $<$900  \\ 
PitsB & $<$1                 &  $<$9           & $<$500  \\  
PitsC & $<$1                &   $<$3           & $<$200 \\ 
PitsD & $<$1                &   $<$2           & $<$150  \\  
PitsE & $<$1                &   $<$7           & $<$400  \\ 
PitsF & $<$1                &   $<$8           & $<$400 \\ 

\end{tabular}
\end{center} 
\vspace{-0.75cm}
\end{table}

Lastly, we would like to make a point that, complex and time-costly model like LDADE or other $K$ values of $25, 50, 100$ is not needed. We can use $K=10$ as the optimal number of features to build a simple FFT model. Hence,

\begin{lesson1}
Simpler method (LDA+FFT) offers similar performance as the sophisticated method (TFIDF+SVM) and the SBSE method (LDADE+SVM). Though simpler LDA+FFT method takes an extra 10  minutes than the baseline, but it is orders of magnitude faster than the SBSE method.
\end{lesson1}




\begin{figure*}[!t]
 {\footnotesize

\begin{tabular}{p{\linewidth}}\hline

\begin{minipage}{.45\linewidth}
\begin{tabular}{p{.95\linewidth}}
PITS\_A Dataset:\\
  \qquad if  \qquad \ \ topic 1 $>$  0.80   \qquad \qquad  then false \\
  \qquad  else if \ \ topic 7 $>$  0.60    \qquad \qquad   then true \\
   \qquad else if \ \ topic 3 $>$  0.65   \qquad \qquad  then true \\
   \qquad else if \ \ topic 5 $\leq$ 0.50    \qquad \qquad  then true \\
   \qquad else \qquad \qquad \qquad \qquad \qquad \qquad  \ false 
\end{tabular}
\end{minipage}
\begin{minipage}{.55\linewidth}
\begin{lstlisting}[
  mathescape,
  columns=fullflexible,
  basicstyle=\fontfamily{times}\selectfont,
]    
     Topic 1: type data line code statu packet word function
     Topic 7: mode point control project attitud rate error prd 
     Topic 3: messag unsign bit code file byte word ptr 
     Topic 5: file variabl code symbol messag line initi access 
            
  \end{lstlisting}
  \end{minipage}
  
  \end{tabular}
  \vspace{1mm}
 \begin{tabular}{p{\linewidth}}\hline
 
 \begin{minipage}{.45\linewidth}
 \begin{tabular}{p{.95\linewidth}}
PITS\_B Dataset:\\
   \qquad if  \qquad  \ \  topic 2 $>$  0.70    \qquad \qquad   then true\\
   \qquad else if \ \ topic 4 $>$  0.75   \qquad \qquad  then false\\
   \qquad else if \ \ topic 7 $>$  0.65   \qquad \qquad  then true\\
   \qquad else if \ \ topic 6 $\leq$ 0.80    \qquad \qquad  then true\\
   \qquad else \qquad \qquad \qquad \qquad \qquad \qquad   \ false  
\end{tabular}
 
\end{minipage}
\begin{minipage}{.55\linewidth}
\begin{lstlisting}[
  mathescape,
  columns=fullflexible,
  basicstyle=\fontfamily{times}\selectfont,
]    
     Topic 2: command gce counter step bgi test state antenna 
     Topic 4: line code function file declar comment return use 
     Topic 7: ace command fsw shall level state trace packet 
     Topic 6: test interfac plan file dmr document section data 
            
  \end{lstlisting}
  \end{minipage}
   \end{tabular}
   
  \vspace{1mm}
   \begin{tabular}{p{\linewidth}}\hline
  \begin{minipage}{.45\linewidth}
   \begin{tabular}{p{.95\linewidth}}
PITS\_C Dataset:\\
  \qquad if  \qquad \ \   topic 1 $>$  0.70    \qquad \qquad  then false \\
  \qquad  else if \ \ topic 6 $>$  0.55   \qquad \qquad  then true \\
  \qquad  else if \ \ topic 8 $>$  0.73   \qquad \qquad   then true \\
   \qquad else if \ \ topic 2 $>$  0.85   \qquad \qquad   then false \\
   \qquad else \qquad \qquad \qquad \qquad \qquad \qquad  \ false 
\end{tabular}
\end{minipage}
\begin{minipage}{.55\linewidth}

\begin{lstlisting}[
  mathescape,
  columns=fullflexible,
  basicstyle=\fontfamily{times}\selectfont,
]    
     Topic 1: requir fsw command specif state specifi shall ground 
     Topic 6: tim trace section document traceabl matrix rqt requir 
     Topic 8: appropri thermal field integr test valid ram violat 
     Topic 2: header zero posit network indic action spacecraft base 
               
  \end{lstlisting}
  \end{minipage}
 
  \end{tabular}
    \vspace{1mm}
  \begin{tabular}{p{\linewidth}}\hline
  \begin{minipage}{.45\linewidth}
  \begin{tabular}{p{.95\linewidth}}
PITS\_D Dataset:\\
  \qquad if  \qquad \ \ topic 6 $\leq$ 0.50    \qquad \qquad  then false \\
  \qquad  else if \ \ topic 1 $>$  0.80    \qquad \qquad   then true \\
  \qquad  else if \ \ topic 4 $>$  0.85   \qquad \qquad   then false \\
   \qquad else if \ \ topic 9 $>$  0.60    \qquad \qquad   then false \\
   \qquad else \qquad \qquad \qquad \qquad \qquad \qquad  \ true 
\end{tabular}
  \end{minipage}
  \vspace{1mm}
\begin{minipage}{.55\linewidth}
\begin{lstlisting}[
  mathescape,
  columns=fullflexible,
  basicstyle=\fontfamily{times}\selectfont,
]    
     Topic 6: essenti record heater occurr indic includ rollov  
     Topic 1: fsw csc trace data field fpa tabl command 
     Topic 4: enabl wheel use disabl respons control protect fault 
     Topic 9: line cpp case switch default projectd file fsw 
                
  \end{lstlisting}
  \end{minipage}
   \end{tabular}
     \vspace{1mm}
  \begin{tabular}{p{\linewidth}} \hline
  \begin{minipage}{.45\linewidth}
\begin{tabular}{p{.95\linewidth}}
PITS\_E Dataset:\\
  \qquad if  \qquad \ \  topic 8  $>$ 0.75   \qquad \qquad  then true \\
   \qquad else if \ \ topic 5  $>$ 0.70    \qquad \qquad   then false\\
  \qquad  else if \ \ topic 7  $>$ 0.50    \qquad \qquad   then false \\
  \qquad  else if \ \ topic 10 $<$ 0.9    \qquad \qquad  then false \\
  \qquad  else \qquad \qquad \qquad \qquad \qquad \qquad  \ true \\
\end{tabular}
\end{minipage}
\begin{minipage}{.55\linewidth}
\begin{lstlisting}[
  mathescape,
  columns=fullflexible,
  basicstyle=\fontfamily{times}\selectfont,
]    
     Topic 8: line file function cmd paramet ccu fsw vml 
     Topic 5: inst phx test project set document softwar verifi 
     Topic 7: ptr size time prioriti ega defin data null 
     Topic 10: word fsw enabl capabl follow vagu present emic 
            
  \end{lstlisting}
  \end{minipage}
 \end{tabular}
 
  \begin{tabular}{p{\linewidth}}\hline
  \begin{minipage}{.45\linewidth}
\begin{tabular}{p{.95\linewidth}}
PITS\_F Dataset:\\
  \qquad if  \qquad \ \    topic 5 $\leq$ 0.80     \qquad \qquad   then false \\
  \qquad  else if \ \ topic 8  $>$ 0.75   \qquad \qquad   then true \\
  \qquad  else if \ \ topic 2  $>$ 0.50    \qquad \qquad  then true \\
  \qquad  else if \ \ topic 9  $>$ 0.65   \qquad \qquad     then true \\
  \qquad  else \qquad \qquad \qquad \qquad \qquad \qquad  \ false \\ 
\end{tabular}
\end{minipage}
\begin{minipage}{.55\linewidth}

\begin{lstlisting}[
  mathescape,
  columns=fullflexible,
  basicstyle=\fontfamily{times}\selectfont,
]  
     Topic 5: requir projectf tabl ref boot bsw fsw section 
     Topic 8: fsw requir test projectf procedur suffici softwar 
     Topic 2: code variabl test point build defin float valu 
     Topic 9: number byte word limit buffer dump ffp error 
               
  \end{lstlisting}
  \end{minipage}
  \end{tabular}
}
\caption{Comprehensible models generated by FFT for all 6 datasets }
\label{fig:comprehensible}
\vspace{-0.5cm}
\end{figure*} 

\textbf{RQ2:  {Is simpler method more explainable or comprehensible against the most common sophisticated and recent state-of-the-art SBSE methods?} }


Beside the comparable performance of the simpler method against the most common sophisticated method and the recent SBSE method, it would not bring any merits to practice for software analytics without having explainable insights that can be easily interpreted from the model. Representative  characteristics  that  help  a  model more  explainable, includes small architecture, easily visualized, and comprise of fewer rules  that  can  quickly  lead  to  decisions. From Table~\ref{tb:dataset}, with large features size range of 550-2000 features from the six datasets of our study, the classifier built on top of that will be too big and complex. Since  2013,  researchers  have  started  focusing  on  using  LDA features instead of TFIDF to offer the comprehensible aspect of the models. However, LDA features only provide better sense of interpretability if we have 10s of features not 100s.  Researchers have showed both the top key words from TFIDF or LDA~\cite{layman2016topic,agrawal2018wrong,xia2015automatic,kochhar2014automatic,xia2014automated} features in an attempt to compensate for the comprehensibility of the model but there were no simple decision-making process embedded with it, so the model is not actionable.

For this study, support vector machines were picked as the most common sophisticated method in text mining. SVMs achieve the results after synthesizing new dimensions through the kernel function which are totally unfamiliar to human users. Hence, it is hard to explain to the users. 

The proposed simple model of FFT with $K=10$ LDA topics, depth $d= 4$, references the trend of only 4 topics from LDA. At each level $d$ of the FFT tree, the existing branch can select for the severeness target, i.e., true (denoted ``1''), or the non-severeness target, i.e., false (denoted ``0''), as it's exiting policies. The exiting policies selected by FFT are a trace of the model sampling around the space toward the sections of the data containing the targets of severe class of bug reports. With this architecture, the LDA+FFT would be more explainable for text mining to determine the severity of the bug. 

Figure~\ref{fig:comprehensible} demonstrates how our models can be explainable. The right hand side  of the figure shows the four most important topics as a list of top relevant words per dataset. 
The left hand side includes decision rules of the best performing FFT tree that fit with the LDA generated topics. Some of the possible interpretations of the FFT models from Figure~\ref{fig:comprehensible} include:

\bi
    \item The FFT tree from PitsC dataset, say for depth 1, the exiting policy says that when a report of the dataset will have probability of topic 1 higher than $0.7$ then that report will be a non-severe report.
    \item In other case, the exiting policies for PitsE FFT  is ``10001''. It starts off with deciding the severeness targeting some low hanging fruit of severe bug reports. Only after clearing away all the non-severe examples at levels two, three, four,  it makes a final ``true'' conclusion. Note that all the exits, except the first and the last, are ``false''.
    \item For PitsF FFT's exiting policies of ``01110''. It is similar to ``10001'' where ``01110'' starts off with clearing away the non-severe examples then commit on finding the target classes and then clear the rest of non-severe examples. Note that all the exits, except the first and the last, are ``true''.
\ei

In practice, business users/experts can use this explainable and comprehensible method to identify a new unseen/not labeled report into severe and non-severe, reducing the time and cost spent by business in labeling these reports~\cite{deng2014scalable,chen2017replicating}.
For e.g., once FFT tree is built on the seen examples using LDA, a new bug report instance will use LDA to automatically come up with topic probabilities of this report (like topic 1 = 0.7, topic 2 = 0.02 and so on). We can then use the probabilities to traverse through the built FFT tree to classify the severeness of the bug report automatically.
With the comparable performance demonstrated in \textbf{\textit{RQ1}}, this method shall confidently give those experts an actionable and intuitive but more scientific way to quickly label the severeness of the bug report. 

Moreover, comprehensibility aspect of the model also let the expert testing theories appropriately. 
For instance, some of the top words from topic 6 generated for the PitsB dataset (Figure~\ref{fig:comprehensible}) include ``\textbf{test, plan, document, data}'' in which test planning topic can be easily inferred from. By following the respective FFT model, the development team would now take test planning into more serious consideration in the software development lifecycle to minimize future sever bugs in the software. The team will have the autonomy to easily refine the method accordingly or generalize this method for future applications, which is the two strongly suggested characteristics of the power of comprehensibility by Gleicher \cite{gleicher2016framework}.


On the other hand, the models generated from complex or SBSE method will look like Figure~\ref{fig:svm_model}. As discussed earlier in \tion{comprehensibility}, SVM model generates synthetic feature space and an imaginary hyperplane boundary that lack the power of explainability of such a model to humans. We can not use such a decision space to reason from or make it actionable.

Altogether, our proposed LDA+FFT method has more actionable and comprehensible aspects against TFIDF+SVM, our most sophisticated method, and LDADE+SVM, SBSE method. Moreover, the cost of running LDA+FFT in \textbf{\textit{RQ1}} will be compensated with the interpretability of the model. Hence,    

\begin{lesson1}
FFT generates fewer rules referencing only 4 topics found by LDA are far more comprehensible than the most common sophisticated  and  SBSE methods.
\end{lesson1}

\section{Discussion}

We found that FFT with small feature space (10 features) found by LDA works as well as SVM with 100s to 1000s TFIDF features and much better than the combination of LDADE and SVM which makes the discussion important on why FFT works. There could be two reasons behind this:
\be
    \item The exit policies selected by FFTs are like a trace of the reasoning jumping around the data. For example, a tree with 11110 policy jumps   towards sections of the data
 containing most severe reports. Also, a 00001 tree shows another model trying to jump away from severe reports  until, in its last step, it does one final jump towards  severe. This tells us that software data could be ``lumpy'', i.e., it divides
into a few separate   regions, each with different properties. 
In such a ``lumpy'' space,
a learning policy like FFT works well since its exit policies let a learner discover how to best
jump between the ``lumps'' and other learners  fail in this coarse-grained  lumpy space~\cite{chen2018applications,fu2018building}.

    \item FFT combines good and bad attributes together to find the best decision policy~\cite{phillips2017fftrees}. FFT finds a rule by identifying the exit policy that has the highest probability of that rule leading to a particular class even if the rule contains mixed class distribution. On the other hand, learners like SVM, transform the data into different feature space which could still contain noisy relationship between the transformed space and the decisions. 
\ee

\noindent
Based on the above discussion, we will need to extend the usage of FFT in other software analytics tasks on more complex data to see whether the results from this paper holds true for them or not.

\section{Threats to Validity}
\label{sect:validity}

As with any empirical study, biases can affect the final
results. Therefore, any conclusions made from this work must consider the following issues in mind.

\textbf{\textit{Order bias}}: With each dataset how data samples are distributed in training and testing set is completely random. Though there could be times when all good samples are binned into training set. To mitigate this order bias, we run
the experiment 25 times by randomly changing the order of the data samples each time.

\textbf{\textit{Sampling bias}} threatens any classification experiment, i.e., what matters here may not be true there. For e.g., the datasets used here comes from the SEACRAFT repository and were supplied by one individual. These datasets have been used in various case studies by various researchers~\cite{layman2016topic,menzies2008automated,menzies2008improving}, i.e., our results are not more biased than many other studies in this arena.
That said, our 6 open-source datasets   are mostly from NASA. Hence
it is an open issue if our results will hold true for both
 proprietary and open source projects from other sources. Also, our FFT results can also be affected by the size of each datasets. These datasets are smaller in corpus size, so in future, we plan to extend this analysis on larger and higher dimensional datasets.

\textbf{\textit{Learner bias}}: For LDADE, we selected  parameters as default as provided by Agrawal et al.~\cite{agrawal2018wrong}. But there could be some datasets where by tuning them there could be larger improvement. We only used SVM as classifier but there could be other classifiers which can change our conclusions. Data Mining is a large and active field and any single study can only use a small subset of the known data miners.

\textbf{\textit{Evaluation bias}}: This paper uses topic similarity ($\Re_n$) for LDADE, and precision and recall for classifiers, but there are other measures which are used in software engineering which
includes perplexity, accuracy, etc. Moreover, based on our experiment, we picked precision, and there would be loss in recall performance and vice-versa. Assessing
the performance of both the metrics together showing there trade-offs is left for future work. 


We would also like to point out that FFTs are only for binary classification, however for multi-class the FFTs can be improvised upon to accommodate this request. 
Also, FFTs do not scale well with 1000s of features and becomes computationally expensive, which can further be improved. 
In this study, we  used a default depth of 4 to build the trees (in total 16 trees are build to find the best one), but we also need to try with other depth size to see what performance changes will we see making it a clear focus for future.

\section{Conclusion}
\label{sect:conclusion}

This paper has shown that a simple and comprehensible data mining algorithm, 
called Fast and Frugal trees (FFTs) developed
by psychological scientist, is remarkably effective for creating few decision rules that are actionable and browsable.

Despite their  succinctness, LDA+FFTs are remarkably effective in showing comparable performance on recall
and precision when compared against the most common technique of TFIDF with SVM as well as state-of-the-art SBSE method (LDADE+SVM).
It can also be said that, we do not need computationally expensive methods to find succinct models.

From the above, we conclude that, there is much for software analytics community that could be learned from psychological science. 
Proponents of complex methods should always baseline against simpler alternative methods.
For example, FFTs could be used as a standard baseline learner against which other software  analytics tools can compare.

\balance
\bibliographystyle{IEEEtran}


\end{document}